\documentclass{article}
\usepackage{caption}
\usepackage{subcaption}
\usepackage{spconf,amsmath,graphicx}
\usepackage{amsmath} 
\usepackage{algorithm} 
\usepackage{algpseudocode}
\usepackage{algorithm,algpseudocode}
\usepackage{graphicx}
\usepackage{caption}
\usepackage{url}

\usepackage{amsbsy}

\usepackage{color,soul} 
\sethlcolor{yellow}

\title{Identifying influential pandemic regions using graph signal variation}
\name{Sudeepini Darapu$^{\ast }$, Subrata Ghosh$^{\ast }$, Abhishek Senapati $^{\dag}$, Chittaranjan Hens$^{\ast }$, Santosh Nannuru$^{\ast }$ \thanks{S.D is supported by research fellowship from IIIT Hyderabad I-HUB, C.H. is supported by the DST-INSPIRE-Faculty grant (Grant No. IFA17-PH193), S.N. is supported by the DST-INSPIRE-Faculty Award (Grant No. IFA17-ENG236)}}
\address{ $^{\ast}$ International Institute of Information Technology, Hyderabad, India\\
$^{\dag}$ Center for Advanced Systems Understanding, Görlitz, Germany}
\begin{document}
%\ninept
%
\setlength{\abovedisplayskip}{3pt}
\setlength{\belowdisplayskip}{3pt}
\maketitle
\begin{abstract}
Developing methods to analyse infection spread is an important step in the study of pandemic and containing them. The principal mode for geographical spreading of pandemics is the movement of population across regions. We are interested in identifying regions (cities, states, or countries) which are influential in aggressively spreading the disease to neighboring regions. We consider a meta-population network with SIR (Susceptible-Infected-Recovered) dynamics and develop graph signal-based metrics to identify influential regions. Specifically, a local variation and a temporal local variation metric is proposed. Simulations indicate usefulness of the local variation metrics over the global graph-based processing such as filtering.
\end{abstract}
\section{Introduction}
\label{sec:intro}

For recent epidemics such as SARS \cite{colizza2007predictability}, and COVID-19 \cite{arenas2020modeling,kraemer2020effect}, the time scale to disseminate the disease from one country to another is a few months with mobility playing a crucial role. 
%It is possible to ignore the movement of diseased people from one patch to another in a well-mixed, homogeneous population, but this is not always the case. 
This motivates us to take into account the reaction-diffusion dynamics \cite{colizza2007predictability,keeling2008modeling,brockmann2013hidden,hens2019spatiotemporal,belik2011natural}, in which the agents interact ("react") within a community and "diffuse" (mobility) in short time scales. %We study a meta-population model captured by multiple regions with interconnecting links representing the movement of the agents across regions, and the interaction mechanism within a region governed by disease dynamics.
We investigate a meta-population model that is represented by a number of interconnected regions, where the links stand for the agent's trans-regional migration and the interaction mechanisms within a region controlled by disease dynamics.

Signal processing techniques are well understood for time and space domain signals but are ineffective for signals over the graph (network) domain. Recently, graph signal processing (GSP) \cite{ortega2018graph,shuman2013emerging} framework has emerged which processes signals \cite{sandryhaila2013discretegft, sandryhaila2014discrete,sandryhaila2013discrete} while being cognizant of the relations between various components of the signal captured by the graph. We analyse the temporally evolving infection data from meta-population network with SIR (Susceptible-Infected-Recovered) dynamics using GSP framework.

Our work focuses on identifying influential nodes defined as the nodes which significantly influence the evolution of the graph signal across the network over time. A few methods have recently been proposed in the literature for this. Graph frequency analysis has been used \cite{li2021graph} to study spread of COVID-19 in US counties. In \cite{pena2016source}, authors process ``John Snow’s Cholera Data" and model cholera transmission as heat diffusion process to localize the source of infection. Using the spectral graph wavelet transform, spatio-temporal patterns of COVID-19 virus spread in Massachusetts are analysed \cite{geng2022analysis}.
%Our work identifies infected source nodes, investigates the spread pattern by identifying influential spreader nodes, and provides detailed spatio-temporal analysis of disease spread patterns using GSP.
%and the conclusions support the assuring prospect of applying GSP tools for epidemiology understanding in graph settings. 

We focus on the propagation pattern of disease in a network of regions (nodes) when regional temporal data is given. Local variation based measures are proposed that can detect regions with weak and strong (influential spreaders) ability to spread the infection to its neighbors. We further show that the proposed measure can be useful for the detection of a secondary weak perturbation. If the force of infection of a particular node is slightly enhanced (which may not be visible from the raw data) during the propagation, the proposed algorithm can detect such an anomaly in an efficient way. 

%For simulation purposes, we are focusing on a meta-population network of the SIR (Susceptible-Infected-Recovered) compartmental model to understand how the disease spread through the patches of populations. 
%The geographical disease spread can be mainly attributed to population movement. With appropriate mapping of geographical locations as nodes and edges connecting them, a graph can compactly capture and model this spread. These graphs and the available information on nodes, jointly termed as graph signals , provide a novel perspective for data representation and analysis.   GSP provides novel ways to process graph signals which incorporates the structural relations present in the signal. 

\section{Graph signal variation}
\vspace{-0.2cm}
Let $\mathcal{G} = (\mathcal{V}, \mathcal{E}, W)$ be an undirected graph with node set $\mathcal{V}=(1,2,\cdots, N)$, edge set $\mathcal{E}$, and symmetric weighted adjacency matrix $W$. The graph Laplacian is defined as $L = D - W$, where $D$ is the diagonal degree matrix.
%whose diagonal element $D_{mm}$ is the aggregate of weights of all edges connected to node $m$. 
Denote the time-varying graph signal as $X = [\mathbf{x}_1,\mathbf{x}_2,\cdots,\mathbf{x}_T]$, where $\mathbf{x}_t=[x_1,x_2,\cdots,x_N]^T$ is the graph signal at time $t$. We now discuss some techniques for capturing graph signal variation. First, a brief review of graph filtering is given. We then extend the notion of total variation \cite{hosseinalipour2017detection} computed on the global graph to local variation computed locally at each node.
\vspace{-0.4cm}

\subsection{Graph high pass filtering}
\vspace{-0.1cm}
The graph Fourier transform (GFT) \cite{sandryhaila2013discretegft} determines how a graph signal varies according to the graph topology by decomposing it into orthonormal components. Given the eigen-decomposition $L=U \Lambda U^T$ of the graph Laplacian $L$, the GFT of a graph signal $\mathbf{x}$ is $\Tilde{\mathbf{x}}=U^T \mathbf{x}$.
%The orthonormal matrix ${U}$ represents the eigenvectors which acts as frequency basis, and a diagonal matrix ${\Lambda}$ consists of non-negative eigenvalues which represents graph frequencies of graph Laplacian.
Let $\mathbf{h}_{\text{HPF}}$ be a binary vector indicating positions of highest $M < N$ graph frequencies. The high pass filtered graph signal \cite{li2021graph} is given by $\mathbf{x}_{\text{HPF}} = H_{\text{HPF}} \mathbf{x}$, where the high pass filter is $H_{\text{HPF}} = U \Tilde{H}_{\text{HPF}} U^T$ and $\Tilde{H}_{\text{HPF}} = \text{diag}(\Tilde{\mathbf{h}}_{\text{HPF}})$. The high pass filter extracts the graph signal component $\mathbf{x}_{\text{HPF}}$ that varies rapidly over the graph $G$. As an alternative to HPF, simple low-complexity approaches for analyzing graph signal variation locally on a node are discussed next.

\vspace{-0.5cm}
\subsection{Total Variation (TV) and Local Variation (LV)}
The total variation of a graph signal $\mathbf{x}_{t}$ is given by
\begin{equation}\label{eq2}
\mathcal{TV} = \sum_{i=1}^{N} \sum_{j\in \mathcal{N}_i}{{(\mathbf{x}_{t}(i)-\mathbf{x}_{t}(j))}^2 W_{ij}}
\end{equation}
where $\mathcal{N}_i$ denotes the set of neighbors of node $i$. A signal's total variation over graph indicates how much it changes globally over $G$.
In order to measure how much a graph signal at a node $i$ varies with its neighbors $\mathcal{N}_i$, we define (per node) local variation,
\begin{equation}\label{LV}
\mathcal{LV}(i,t) = \sum_{j\in \mathcal{N}_i}{{(\mathbf{x}_{t}(i,t)-\mathbf{x}_{t}(j,t))}^2 W_{ij}}.
\end{equation}
Large $\mathcal{LV}(i,t)$ values imply the signal at node $i$ and time $t$ is highly varying compared to its neighbors, whereas small $\mathcal{LV}(i,t)$ implies the signal at node $i$ does not vary with its neighbors. It can be used to identify nodes with locally anomalous graph signal variation. Specifically, for epidemic analysis over networks, we use local variation to identify influential nodes which contribute more to the disease spread. HPF-based processing captures the signal variation over the global network. In contrast, LV-based processing captures the variation locally at each node.
\vspace{-0.5cm}

\subsection{Temporal Local Variation (TLV)}
The high pass filtering and local variation capture signal dynamics of the vertex with respect to its neighborhood (global and local). For temporally evolving graph signals, we can extend the notion of local variation into temporal dimension and define temporal local variation as follows,
\begin{equation}\label{TLV}
\mathcal{LV}_{\mathcal{T}}(i,t)=
     %\overbrace{
      \underbrace{{{(\mathbf{x}_{t}(i)-\mathbf{x}_{t-1}(i))}^2}}_\text{Temporal } 
      %\underbrace{\mathcal{LV}(i)}_\text{Spatial}
      \overbrace{\sum_{j\in \mathcal{N}_i}{{(\mathbf{x}_{t}(i)-\mathbf{x}_{t}(j))}^2 W_{ij}}.}^\text{Spatial}
     %}^\text{TLV}
\end{equation}
The TLV, which is product of temporal variation and local variation, captures the simultaneous variations of the graph signal in vertex and time. For epidemic analysis, the TLV can identify anomalous nodes which are temporally active.

\begin{figure}[t]
    \centering
    \includegraphics[ width=\linewidth]{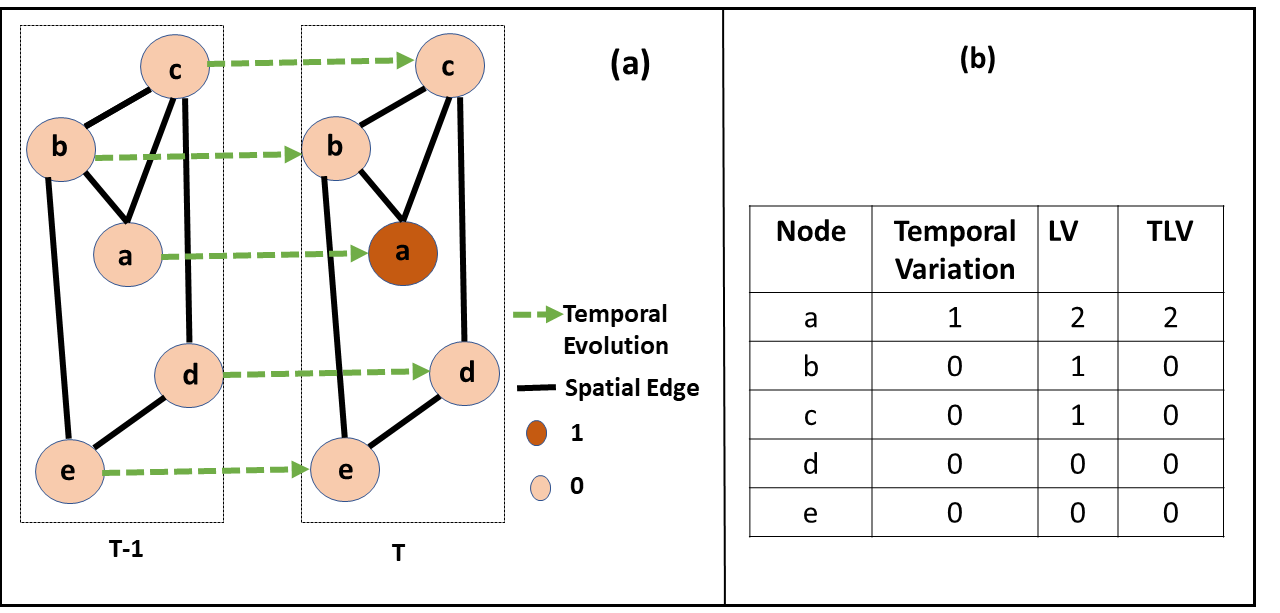} 
    \caption{Illustration of LV and TLV using a simple  graph} \label{fig:toy graph}
\end{figure}

A demonstration of LV and TLV is shown in Fig \ref{fig:toy graph} using a five node graph, $\mathcal{V}=[a,b,c,d,e]$. The evolving graph signal (color coded) is shown in Fig \ref{fig:toy graph}(a) at times $T-1$ and $T$. Dashed arrows indicate the evolution of the graph signal in time. The temporal variation \eqref{TLV}, local variation \eqref{LV} and temporal local variation \eqref{TLV} are computed at time $T$ and shown in the table in Fig \ref{fig:toy graph}(b). The nodes $a, b$ and $c$ have non-zero $\mathcal{LV}$ reflecting the signal structure in the local neighbourhood. Since nodes $b$ and $c$ do not show temporal variation, their TLV is zero. Only node $a$ shows simultaneous variation in time and in its local neighbourhood giving it a non-zero TLV.
%%
%By computing LV we can see how node $a$ influences its connected neighbours $b,c$ in \ref{fig:toy graph}b. It thus captures only spatial variation. It is also necessary to include temporal information when identifying nodes that vary spatially and temporally. As a result, we developed a metric that captures both spatial and temporal variation, i.e., TLV. The table in \ref{fig:toy graph}b shows that TLV identifies spatially and temporally varying nodes. Thus it is possible to detect infected source nodes and influential spreader nodes that propagate graph signals spatiotemporally using TLV.

\section{Epidemic analysis using GSP}
\label{sec:chapter3}
%In this section, a brief overview of graph construction, spatial-temporal graph signal generation using the SIR model, and methods for identifying infected source nodes and influential spreader nodes is discussed.
%\vspace{-0.5cm}

%\subsection{Simulation data generation using SIR model}
\subsection{Network SIR model}
\label{SEC:data generation}
For illustration, we consider the standard SIR (susceptible-infected-recovered) model to depict how an epidemic propagates through a population. In SIR dynamics \cite{keeling2008modeling}, a set of 3-coupled equations describe how the disease spreads within a location: %into patches or locations:
\begin{eqnarray}
%\begin{array}{llll}
\label{eq:model} 
\dot{S}(t) &=& \displaystyle  -\beta \frac{S(t) I(t)}{H} ,\\
\dot{I}(t)  &=& \displaystyle \beta \frac{S(t) I(t)}{H} - \gamma I(t), \\   
\dot{R}(t) &=&  \gamma I(t),
%\end{array}  
\end{eqnarray}
where susceptible ($S$), infected ($I$), and recovered ($R$) are the three ``compartments" of the population. The rates of disease transmission and average recovery are $\beta$ and $\gamma$, respectively. $H$ is the population size of that location. We now consider a spatial network of $N$ locations and employ a coupled population (meta-population) network with SIR dynamics.
%Now we employed a coupled population(meta-population) network with SIR dynamics to validate the effectiveness of our approach to capture epidemics spreading in a network. We consider a spatial network of $N$ locations, i.e., $N$ number of communities/nodes. 
Let the $i^{th}$ location contain $H_i (i=1,2,...,N)$ individuals. We formulate the meta-population network equations  \cite{brockmann2013hidden,ghosh2021optimal} by taking into account the dispersion through diffusion of normalized susceptible ($\mathcal{S}_i=\frac{S_i}{H_i}$), infected ($\mathcal{I}_i=\frac{I_i}{H_i}$), and recovered ($\mathcal{R}_i=\frac{R_i}{H_i}$) individuals,
   \begin{eqnarray}
   	\begin{array}{llll}
   		\dot{\mathcal{S}_i} &=& \displaystyle -\beta_{i}  \mathcal{S}_i \mathcal{I}_i  + \frac {\kappa}{d_{i}} \sum_{j\in \mathcal{N}_i} W_{ij} (\mathcal{S}_{j}-\mathcal{S}_{i}), \\
   		%\label{Net_succ_Dynamics}   
   		\dot{\mathcal{I}_i} &=& \displaystyle \beta_{i}  \mathcal{S}_i \mathcal{I}_i -\displaystyle \gamma_{i} \mathcal{I}_i +
   		\frac {\kappa}{d_{i}} \sum_{j\in \mathcal{N}_i} W_{ij} (\mathcal{I}_{j}-\mathcal{I}_{i}).
   	\end{array}
   	\label{eq:network_model}   
   \end{eqnarray}
%Here, $W_{ij}$ is the element of the {binary  adjacency matrix $W$ revealing the connectivity pattern among the patches}. $d_i=\sum_{j=1}^N W_{ij}$ is the degree (number of neighbors) of the $i^{\rm th}$ patch. 
Here $W$ is the weighted adjacency matrix representing the spatial network and $d_i=\sum_{j\in \mathcal{N}_i} W_{ij}$ is the degree of the $i^{\rm th}$ node. The population migration is modeled through the diffusive term $\sum_{j\in \mathcal{N}_i} W_{ij} (X_{j}-X_{i})$ connected through two compartments ${X: \mathcal{S}_i,\mathcal{I}_i}$. Strength of the migration is determined by $\kappa$. To solve the ODE equations, we utilised Runge-Kutta (RK4)\cite{runge1895numerische} method with adaptive step size.  

\subsection{Methodology for identifying influential nodes}
%infection source node detection?
Given a graph $\mathcal{G}$ and the graph signal $X$, we wish to identify the nodes which significantly influence evolution of graph signal in time. We use the proposed graph variation metrics to develop an algorithm to detect these time-varying influential nodes. When applied to epidemic models, these influential nodes can help us understand the evolution of disease. Though we focus on epidemics, the method can potentially be applied to other time varying graph signals.

Let $\mathbf{x}_{t}(i)$ be the infected fraction of the population at node $i$ and time $t$. 
A temporal sliding window of length $\ell$ can be applied on $X$ to obtain a time-averaged graph signal $\mathcal{Y}$ if desired. An effective node classification method should capture the changes in the graph signal at both spatial and temporal scales. We classify the nodes by computing the graph variations on $\mathcal{Y}$ . The pseudo-code of our method is given in Algorithm \ref{ALG1} which takes as inputs the graph signal $X$ and the graph $\mathcal{G} = (\mathcal{V}, \mathcal{E}, W)$.

We first compute the graph variation (LV or TLV) of the input signal (line 4) to obtain $\mathcal{Y_V}$. This step can be replaced with high pass filter (for HPF analysis), $H_{\text{HPF}}$. Then, temporal mean is computed and a threshold $\tau$ is obtained from the mean of top $k\%$ of the nodes. Finally, the nodes are classified at each time step by comparing the graph variation with $\tau$ (lines $10-16$). For raw data ($\mathcal{Y}$) processing, we use lines $7-16$.

%Similarly, calculating $LV$ (\ref{LV}) in step $3$ of Algorithm \ref{ALG1} is another method for identifying the infected source and influential spreader nodes. In order to compare both $LV$ and $TLV$ results, we also carried out a similar analysis using HPF except steps $1-5$ of Algorithm \ref{ALG1} are replaced by vertex domain high-pass filtering using (\ref{HPF}). We generated HPF using the highest $20\%$ of frequencies. We considered magnitude of high pass filtered signal for further processing of steps $6-15$ in Algorithm \ref{ALG1}.

\begin{algorithm}
\caption{ Identifying influential nodes \\
Input: $X$, $W$, $\kappa$}
\begin{algorithmic}[1]
    %$\mathbf{Y}$ = Continuous time weekly average of $\mathbf{X}$
    \State $\mathcal{Y}=$ time-windowed average of $X$
	\For {$ t=1,2,\ldots,T$}
		\For {$i=1,2,\ldots,N$}
			\State Compute $ \mathcal{Y}_V(i,t)$ using \eqref{LV} or \eqref{TLV}
		\EndFor
	\EndFor
	\State Compute average across time:
		$\mathbf{ \mathcal{Y}_{VA}}=mean(\mathcal{Y}_V)$
    \State Sort in descending order:
		$\mathbf{\mathcal{Y}_{VA_Sorted}}=sort(\mathbf{\mathcal{Y}_{VA}})$
    \State Threshold: $\tau = $ mean of top $k\%$ of $\mathbf{\mathcal{Y}_{VA_Sorted}}$
	\For {$ t=1,2,\ldots,T$}
		\For {$i=1,2,\ldots,N$}
	    	\If{ $ \mathcal{Y}_V(i,t)\geq \tau$}
			  \State $j$ is a influential node
			\EndIf
		\EndFor
	\EndFor
\end{algorithmic} 
\label{ALG1}
\end{algorithm}

\section{Simulations and results}

\subsection{Graph construction}
%\vspace{-0.1cm}
A random, distance-based graph is constructed. Uniform random sampling is performed within a $10 \times 10$ square region to obtain $N=600$ graph nodes. Nearby nodes within a distance of $\alpha=1.667$ are connected with edge weight
\begin{equation}\label{eq6}
    W_{ij} =
        \begin{cases}
            e^{\frac{-{\text{dist}(i,j)}^2}{\tau^2}}, \; \text{dist}(i,j) \leq \alpha, i \neq j \\
            0, & i=j.
        \end{cases}
\end{equation}
where $\text{dist}(i,j)$ is the Euclidean distance between nodes $i$ and $j$ and parameter $\tau^2$ is the scaling factor. The generated graph tends to have both locally sparse and locally dense connections (shown in \cite{SDP}). The intuition for considering distance-based graphs is that, in real-world scenarios, if roads play a major role in commute flow between locations, then infectious disease could propagate rapidly from an infected location to its nearest neighboring locations.
%To avoid fully connected graph, we fix a distance threshold $\alpha$ using (\ref{eq7}) such that higher $\sigma$ results in dense connections of edges.
%\begin{equation}\label{eq7}
%    \alpha = \sigma \frac{Plane Dimension}{Nodes}
%\end{equation}
%To test the effectiveness of our algorithm, we used a  simple distance-based graph, static for all time stamps. 

%\begin{figure}[t]
%\centering
%\includegraphics[width=1\linewidth]{graph.png}
%\caption{ Graph}
%\label{fig:graph}
%\end{figure}

\subsection{Epidemic data generation}
\begin{figure}[t]
        \centering
      \includegraphics[width=1\linewidth]{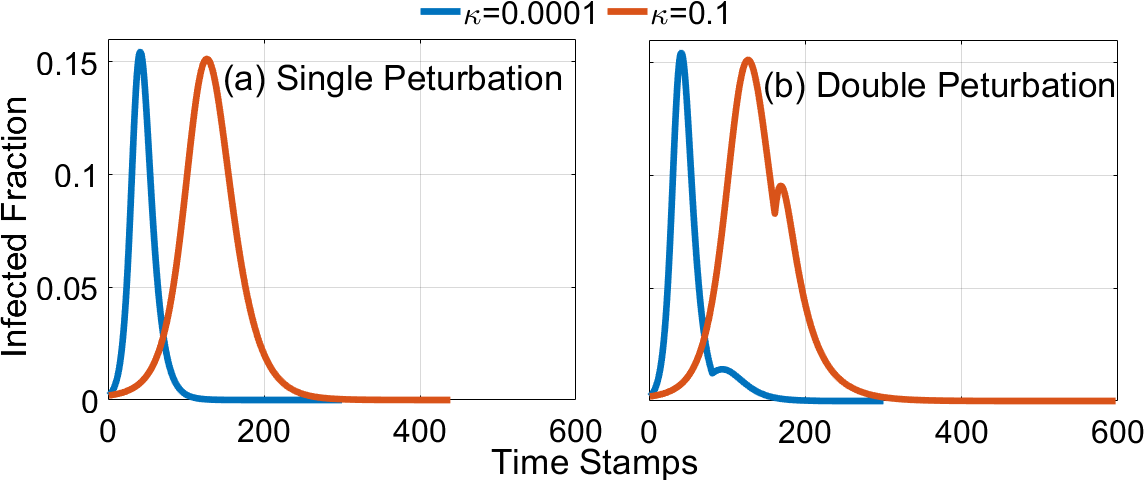} \caption{Single and double perturbation at infected node $588$}\label{fig:fig3s}
\end{figure}

\begin{figure*}[t]
    \centering
    \includegraphics[width=0.44\linewidth]{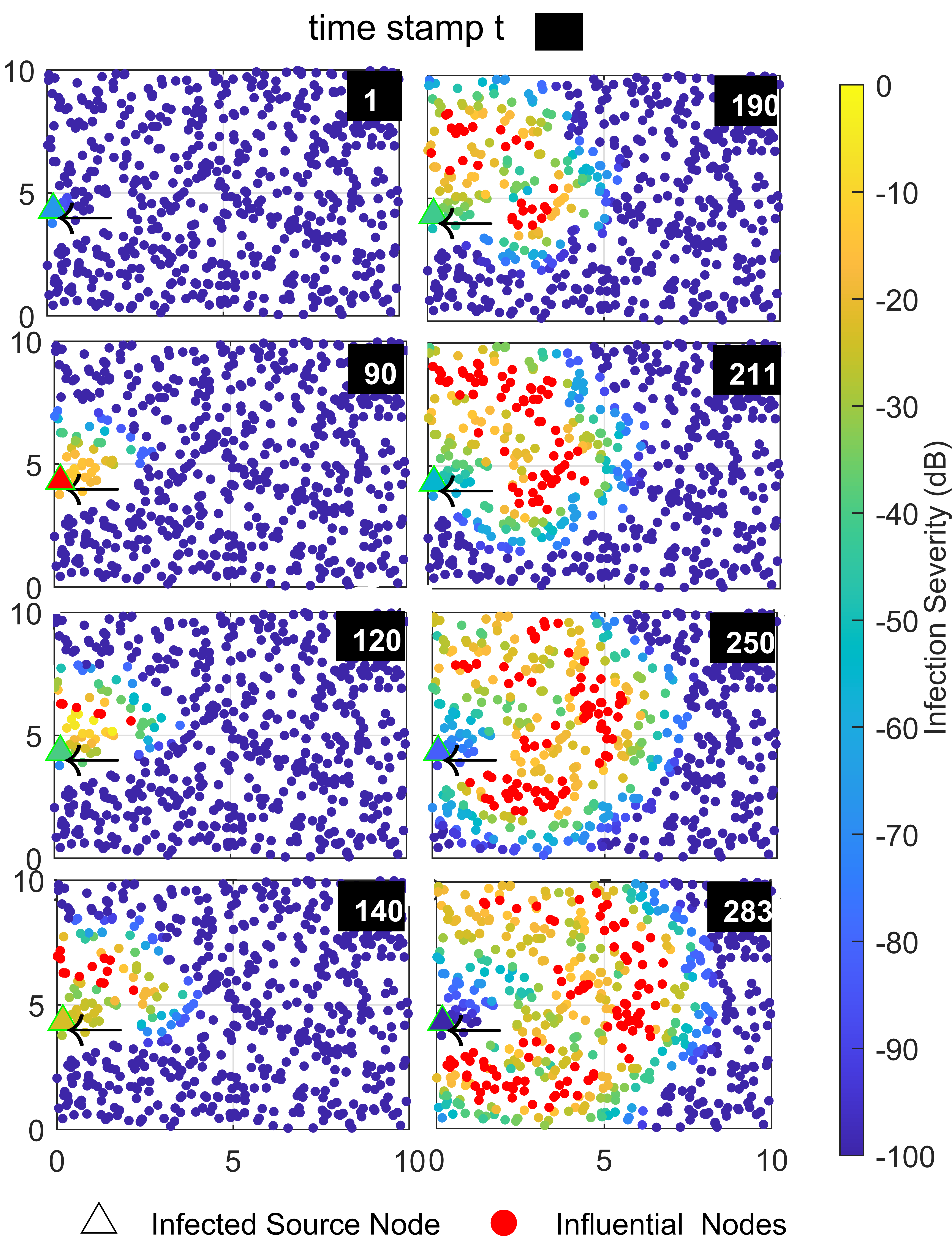} 
    \includegraphics[width=0.465\linewidth]{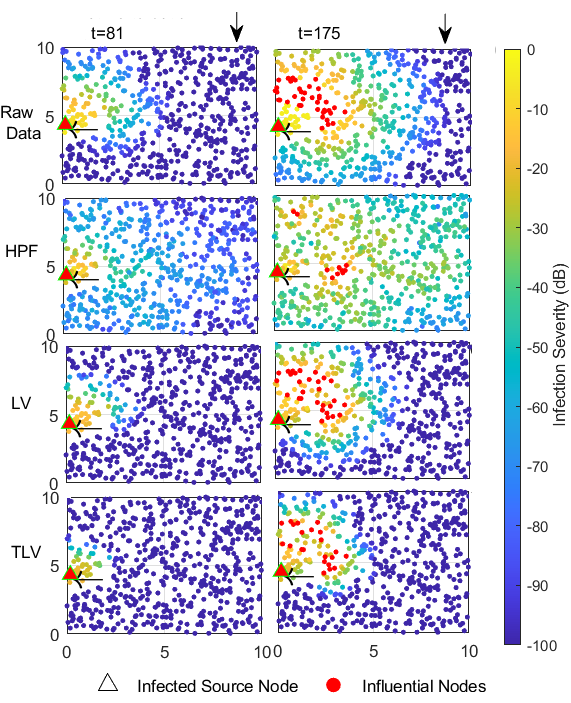}
    \caption{{\bf Left:} Evolution of disease from infected source node of single perturbation for $\kappa=0.1$ using TLV. {\bf Right:} Identification of single and double perturbation at infected node $\# 588$ for $\kappa=0.1$}
    %\label{fig:single}
    %\caption{Single and Double perturbation at infected node $588$}
    \label{fig:double}
\end{figure*}

We consider two scenarios for the creation of synthetic graph signal data $X$ from the network model:\\
\textbf{Single perturbation:} We chose a single node at random (node $\#588$) and introduced an initial infection ($0.2 \%$ of the total population) while keeping the transmission rate ($\beta_i = \beta =0.3$) and recovery ($\gamma_i=\gamma=0.1$) rate same for all the network nodes. The initial $(t = 0)$ infection is zero for all nodes except for node $\#588$. We use two coupling strengths -- low ($\kappa=0.0001$) and high ($\kappa=0.1$), to simulate the epidemic data. Fig \ref{fig:fig3s}(a) shows time evolution of the initially infected node's infection (we simulate $300$ time steps).
\\
%To create the data set for the investigation of the disease propagation pattern in the network from the initially infected node, we used two separate low and high coupling strengths ($k=0.0001, 0.1$).
\textbf{Double perturbation:} The process is the same as above, expect that when node $\#588$ reaches its peak value, a second perturbation is applied to it by altering the value of $\beta$ for a period of time. Specifically, the $\beta$ value is changed from $0.3$ to $0.8$, at $t=161$ when $\kappa=0.1$ ($t=81$ when $\kappa=0.0001$), for a period of $50$ time step. Fig \ref{fig:fig3s}(b) displays the temporal dynamics of infection for this node ($600$ time steps). 
%The second perturbation occurs at $t=81$ when $\kappa=0.0001$, and at $t=161$ when $\kappa=0.1$.
\\
{\bf Data processing and plotting:} Using the graph $G$ and graph signal $X$, we can carry out epidemic analysis using Algorithm \ref{ALG1} with results are presented in Fig \ref{fig:double}. Various stages of epidemic are presented with arrow indicating the infected source node. The identified influential nodes are highlighted in red for visualization. We normalize data to $[0, 1]$ and plot using the logarithmic scale.

%$10 \log_{10}(LV_{GN})$
% \begin{equation}
% \label{eq9}
%  LV_{GN}=\frac{LV_G}{max(max(LV_G))}
%  \end{equation}
%\begin{figure}
        %\centering  \includegraphics[width=\linewidth]{PEAKDIST.png}             \caption{Peak Time vs distance from infected source node}\label{fig:fig single}
%\end{figure}
\vspace{0.5cm}

\subsection{Single perturbation analysis}
Our primary objective here is to understand how disease spreads from a single infected node to the entire network. 
Results of Algorithm \ref{ALG1} when using temporal local variation are shown in Fig \ref{fig:double} (Left) at various time steps. The TLV detects the infected source node and shows how the epidemic spreads from the infected source node over the network. It can capture a variety of graph signal variations as indicated by differently colored nodes.
Disease spread is driven by the influential spreader nodes highlighted in red. Identifying such nodes can help policy makers and health department to take precautionary measures in these regions.
%For high coupling, $k=0.1$, data does not provide insights into the network's spreading behaviour [Fig \ref{fig:fig single}(b)], whereas, a linear relationship between peak time and distance can be found in low coupling strength, $k=0.0001$.
%Therefore, Low coupling can detect how epidemics evolve throughout the network. Using our approach we can also find this relationship in low coupling data. 
%temporally by identifying the influential spreader nodes. 
%Time series plot of the infected source node with high coupling coefficient is shown in Fig \ref{fig:fig3s}b. 
%i.e., it produces various spatial clusters of various frequency bands, such as nodes with smooth variations are clustered in blue and nodes with high variations are clustered in yellow.  
%In addition, the colour of the clustered nodes also indicates the infection severity level and aids in the investigation of the disease spread pattern.

\subsection{Double perturbation analysis}
%The process is the same as in the single perturbation case, with the exception that now we apply a second perturbation for a small time period at the same node after achieving the peak value. 
%The second wave into the network driven by the second perturbation is not captured by model data. Therefore, we demonstrate second wave propagation into the network using our approach. 
Fig \ref{fig:double} (Right) shows the results of raw data, HPF, LV, and TLV on data from double perturbation simulation with $\kappa=0.1$.  
%For comparison we plotted raw data by processing $7-16$ lines of Algorithm \ref{ALG1}.
The methods are able to identify the infection source node $\# 588$ during both the perturbations ($t = 81$ and $t = 175$). 
%The second perturbation is not captured by the raw data.
The HPF based processing does not capture the localized spreading of infection. This is because high frequency are not necessarily localized and show variation over the entire graph. As a result, it is unable to identify influential nodes. Additionally, it is unclear how many frequencies should be used for processing. Detailed plotting results of single and double perturbations for $\kappa=0.1$ and $0.0001$ are shown in \cite{SDP}.

Nodes with spatially varying graph signals are identified by LV, while nodes with spatio-temporally varying graph signals are identified by TLV. A majority of influential nodes detected by LV and TLV are same. Since TLV also accounts for temporal signal evolution, some of the detected nodes are different for LV and TLV. 
Though raw data captures nodes with highest infection, it is unable to capture infection activity (see \cite{SDP} for details).
The computational complexity of Algorithm \ref{ALG1} for LV and TLV is $\mathcal{O}(N^2T)$ whereas for HPF is $\mathcal{O}(N^3+N^2T))$ due to the eigenvalue decomposition.

\vspace{-0.7cm}

\section{Conclusion and future work}
\vspace{-0.2cm}
We examined the spreading pattern of disease from an infected source node using graph signal variation. We introduced low-complexity local variation metrics and devised an algorithm to identify influential nodes based on these metrics. Visualization of local variation and temporal local variation reveal major hot-spots which enable rapid spread of the disease. This can assist policymakers in identifying and classifying regions based on infection severity and follow up with necessary preventive actions. Our further study will focus on analyzing more complicated infection spread simulations and applying our algorithm on real-world data sets of disease. 

\bibliographystyle{IEEEbib}
\bibliography{refs.bib}

\begin{thebibliography}{10}

\bibitem{colizza2007predictability}
V.~Colizza, A.~Barrat, M.~Barth{\'e}lemy, and A.~Vespignani,
\newblock ``Predictability and epidemic pathways in global outbreaks of
  infectious diseases: the sars case study,''
\newblock {\em BMC medicine}, vol. 5, no. 1, pp. 1--13, 2007.

\bibitem{arenas2020modeling}
A.~Arenas, W.~Cota, J.~G{\'o}mez-Garde{\~n}es, S.~G{\'o}mez, C.~Granell, J.~T.
  Matamalas, D.~Soriano-Pa{\~n}os, and B.~Steinegger,
\newblock ``Modeling the spatiotemporal epidemic spreading of covid-19 and the
  impact of mobility and social distancing interventions,''
\newblock {\em Physical Review X}, vol. 10, no. 4, pp. 041055, 2020.

\bibitem{kraemer2020effect}
M.~UG. Kraemer, C.~Yang, B.~Gutierrez, C.~Wu, B.~Klein, D.~M. Pigott, Open
  COVID-19 Data~Working Group†, L.~Du~Plessis, N.~R. Faria, R.~Li, et~al.,
\newblock ``The effect of human mobility and control measures on the covid-19
  epidemic in china,''
\newblock {\em Science}, vol. 368, no. 6490, pp. 493--497, 2020.

\bibitem{keeling2008modeling}
MJ. Keeling and P.~Rohani,
\newblock ``Modeling infectious diseases in humans and animals princeton
  univ,''
\newblock {\em Princeton, NJ}, 2008.

\bibitem{brockmann2013hidden}
D.~Brockmann and D.~Helbing,
\newblock ``The hidden geometry of complex, network-driven contagion
  phenomena,''
\newblock {\em science}, vol. 342, no. 6164, pp. 1337--1342, 2013.

\bibitem{hens2019spatiotemporal}
C.~Hens, U.~Harush, S.~Haber, R.~Cohen, and B.~Barzel,
\newblock ``Spatiotemporal signal propagation in complex networks,''
\newblock {\em Nature Physics}, vol. 15, no. 4, pp. 403--412, 2019.

\bibitem{belik2011natural}
V.~Belik, T.~Geisel, and D.~Brockmann,
\newblock ``Natural human mobility patterns and spatial spread of infectious
  diseases,''
\newblock {\em Physical Review X}, vol. 1, no. 1, pp. 011001, 2011.

\bibitem{ortega2018graph}
A.~Ortega, P.~Frossard, J.~Kova{\v{c}}evi{\'c}, J.MF. Moura, and
  P.~Vandergheynst,
\newblock ``Graph signal processing: Overview, challenges, and applications,''
\newblock {\em Proceedings of the IEEE}, vol. 106, no. 5, pp. 808--828, 2018.

\bibitem{shuman2013emerging}
D.~I. Shuman, S.~K. Narang, P.~Frossard, A.~Ortega, and P.~Vandergheynst,
\newblock ``The emerging field of signal processing on graphs: Extending
  high-dimensional data analysis to networks and other irregular domains,''
\newblock {\em IEEE signal processing magazine}, vol. 30, no. 3, pp. 83--98,
  2013.

\bibitem{sandryhaila2013discretegft}
A.~Sandryhaila and J.~Moura,
\newblock ``Discrete signal processing on graphs: Graph fourier transform,''
\newblock in {\em 2013 IEEE International Conference on Acoustics, Speech and
  Signal Processing}. IEEE, 2013, pp. 6167--6170.

\bibitem{sandryhaila2014discrete}
A.~Sandryhaila and J.~MF. Moura,
\newblock ``Discrete signal processing on graphs: Frequency analysis,''
\newblock {\em IEEE Transactions on Signal Processing}, vol. 62, no. 12, pp.
  3042--3054, 2014.

\bibitem{sandryhaila2013discrete}
A.~Sandryhaila and J.~Moura,
\newblock ``Discrete signal processing on graphs: Graph filters,''
\newblock in {\em 2013 IEEE International Conference on Acoustics, Speech and
  Signal Processing}. IEEE, 2013, pp. 6163--6166.

\bibitem{li2021graph}
Y.~Li and G.~Mateos,
\newblock ``Graph frequency analysis of covid-19 incidence to identify
  county-level contagion patterns in the united states,''
\newblock in {\em ICASSP 2021-2021 IEEE International Conference on Acoustics,
  Speech and Signal Processing (ICASSP)}. IEEE, 2021, pp. 3230--3234.

\bibitem{pena2016source}
R.~Pena, X.~Bresson, and P.~Vandergheynst,
\newblock ``Source localization on graphs via l1 recovery and spectral graph
  theory,''
\newblock in {\em 2016 IEEE 12th Image, Video, and Multidimensional Signal
  Processing Workshop (IVMSP)}. Ieee, 2016, pp. 1--5.

\bibitem{geng2022analysis}
R.~Geng, Y.~Gao, H.~Zhang, and J.~Zu,
\newblock ``Analysis of the spatio-temporal dynamics of covid-19 in
  massachusetts via spectral graph wavelet theory,''
\newblock {\em IEEE Transactions on Signal and Information Processing over
  Networks}, vol. 8, pp. 670--683, 2022.

\bibitem{hosseinalipour2017detection}
S.~Hosseinalipour, J.~Wang, H.~Dai, and W.~Wang,
\newblock ``Detection of infections using graph signal processing in
  heterogeneous networks,''
\newblock in {\em GLOBECOM 2017-2017 IEEE Global Communications Conference}.
  IEEE, 2017, pp. 1--6.

\bibitem{ghosh2021optimal}
S.~Ghosh, A.~Senapati, J.~Chattopadhyay, C.~Hens, and D.~Ghosh,
\newblock ``Optimal test-kit-based intervention strategy of epidemic spreading
  in heterogeneous complex networks,''
\newblock {\em Chaos: An Interdisciplinary Journal of Nonlinear Science}, vol.
  31, no. 7, pp. 071101, 2021.

\bibitem{runge1895numerische}
C.~Runge,
\newblock ``{\"U}ber die numerische aufl{\"o}sung von
  differentialgleichungen,''
\newblock {\em Mathematische Annalen}, vol. 46, no. 2, pp. 167--178, 1895.

\bibitem{SDP}
S.~Darapu, S.~Ghosh, A.~Senapti, C.~Hens, and S.~Nannuru,
\newblock ``Supplementarydocument,''
  \url{https://github.com/DSudeepiniReddy/Icassp2023}, 2022.

\end{thebibliography}

\end{document}